\documentstyle[emulateapj]{article}
\input psfig
\begin{document}
\submitted{}
\title{A Simultaneous Constraint on the Amplitude and Gaussianity of 
Mass Fluctuations in the Universe}
\author{James Robinson\altaffilmark{1}, Eric Gawiser\altaffilmark{2} and Joseph
Silk\altaffilmark{1,2,3}}   
\affil{University of California, Berkeley CA 94720-3411}
\authoremail{jhr@astron.berkeley.edu}
\altaffiltext{1}{Department of Astronomy}
\altaffiltext{2}{Department of Physics}
\altaffiltext{3}{Center for Particle Astrophysics}

\authoremail{jhr@astron.berkeley.edu}

\begin{abstract}

We consider constraints on the amplitude of mass fluctuations in the
universe, $\sigma_8$, derived from two simple observations: the present number
density of clusters and the amplitude of their correlation function. Allowing
for the possibility that the primordial fluctuations are non-gaussian
introduces a degeneracy in the value of $\sigma_8$ 
preferred by each
of these constraints. However, when the constraints are
taken together this degeneracy is broken,
yielding a precise determination of $\sigma_8$ 
and the degree of non-gaussianity for a given background cosmology. 
For a flat, $\Omega_m=1$ universe with a power spectrum
parameterized by a CDM shape parameter $\Gamma=0.2$, we find that the
perturbations are consistent with a gaussian distribution with
$\sigma_8=0.49^{+0.08}_{-0.07}$ (95\% limits).
For some popular choices of background model, including the favored
low matter density models, the
hypothesis that the primordial fluctuations are gaussian is ruled out
with a high degree of confidence.
\end{abstract}
\keywords{Cosmology: observations and theory ---  galaxies:
clusters: general --- 
large-scale
structure of the universe.}

\section{Introduction}

Most studies of structure formation in the universe assume that the
primordial density perturbations are gaussian. Standard inflationary theories
predict gaussian perturbations, and the central limit theorem tells us
that any theory involving the superposition of many random processes
will give rise to approximately gaussian fluctuations. However, 
many well motivated theories predict non-gaussian initial conditions,
including topological
defect theories (Kibble\markcite{k76} 1976) and certain
forms of inflation  (Peebles\markcite{P83,P97} 1983,1997). So far, no
convincing observational evidence has been found to confirm or refute
the gaussian hypothesis.

Clusters of galaxies, however, provide us with a 
unique probe of possible
non-gaussianity. Being the most massive collapsed 
structures, they correspond to rare peaks in
the primordial density field, so their statistics
respond very sensitively to non-gaussianity in  the initial matter
distribution. In addition, an analysis of the formation of clusters 
from given
initial conditions requires primarily gravitational physics, 
and is
largely free of complications from star formation and feedback.
Finally, although the gravitational evolution which must be modeled is
non-linear, well-tested analytical approximation schemes exist
for studying the statistics of the resulting cluster distribution.
 
In this work we adapt these analytic schemes to the
case of non-gaussian fluctuations,
 and use them to make predictions
for two simple observations, 
the number density of clusters and the amplitude
of cluster correlations. We show that
the possibility of
non-gaussianity introduces a degeneracy in the value of $\sigma_8$ 
(where $\sigma_R$
is the {\it rms} overdensity in a sphere of radius $R\;h^{-1}$Mpc)
preferred by
each of these observations. However, combining
the two constraints breaks the degeneracy, allowing
a precise determination of $\sigma_8$ and the degree of
non-gaussianity in the universe.


\section{Model}
\label{method}

We describe the primordial density field by
specifying both the power spectrum $P(k)$ and the probability density
function (PDF) $p_R(\delta)$
of the fractional over-density $\delta$
 after a top-hat smoothing on scale $R$. 
We make the simplifying assumption that
only the {\it rms} level
of fluctuations changes as a function of scale, that
is  
$p_R(\delta)=P(\delta/\sigma_R)/{\sigma_R}$
where $P(y)$ is a rescaled PDF, whose distribution has zero mean and 
an {\it rms} value of
1. This assumption should hold reasonably well 
for topological defect
theories, whose perturbations are scale invariant over a range of
scales, and more generally, will not affect our results provided the
form of the PDF does not change significantly over the narrow range of
scales relevant to cluster formation.

The Press-Schechter (1974) formalism (hereafter PS)
allows us to
compute the number density of clusters within a given
background cosmology for gaussian fluctuations.
The derivation suggests a 
simple way to adapt the PS formula
for theories with a general PDF (as done by Chiu, Ostriker \&
Strauss\markcite{COS}
1997):
The probability that a
given region of space lies inside a collapsed region with
pre-collapse radius larger than $R$ is taken to be 
$\int_{\delta_C}^{\infty} p_R(\delta) d \delta$, where 
$\delta_C$ is a critical threshold for collapse.
Differentiating with respect to $R$ and dividing by the
pre-collapse volume we obtain the number density $n(R)$ of collapsed
objects with pre-collapse radius R, where
\begin{equation}
n(R)dR= \frac{3f}{4\pi R^3} \left| \frac{d}{dR} \left[
\int_{\delta_c}^{\infty} p_R(\delta) d\delta \right] \right| dR.
\end{equation}
As in the gaussian case (Bond et al.\markcite{B91}~1991),
we have multiplied by a correction factor
$f=1/\int_{0}^{\infty}P(y)dy$, to ensure that the entire
mass of the universe is accounted for. 
We integrate the above expression to obtain cumulative number
counts $N_{>R}$ for all objects with a pre-collapse radius greater
than $R$:
\begin{eqnarray}
\label{eqn-n}
N_{>R}
&=& \int_R^\infty dR' \frac{3f}{4\pi R'^3} P[y_c(R')] \frac{dy_c(R')}{dR'}
\end{eqnarray}
where $y_c(R) = \delta_c/\sigma_R$. We convert to cumulative number
counts as a function of mass and temperature using the method of
Eke, Cole \& Frenk\markcite{ECF} (1996).

The peak-background split, first suggested by
Kaiser\markcite{K84} (1984), provides
a simple scheme for  
estimating the bias, and hence the correlation amplitude,
of collapsed objects. Consider the manner in which the
addition of a long wavelength background
fluctuation $\delta_b({\bf x})$ modulates the local number density of
collapsed objects. The threshold for critical collapse due to short
wavelength fluctuations
is modified
locally from $\delta_c$ to $\delta_c-\delta_b({\bf x})$, so the number
density of collapsed objects is modified from its background
value $N_{>R}$ to 
$N_{>R}({\bf x})=(1+\delta_b({\bf x}))N_{>R}|_{\delta_c\rightarrow
\delta_c-\delta_b({\bf x})}$. The factor 
$1+\delta_b({\bf x})$ comes from the linear growth of perturbations. Taylor
expanding to first order in $\delta_b$ we obtain
\begin{equation}
N_{>R} ({\bf x}) = N_{>R} (1+\delta_b({\bf
 x}))
 -\delta_b({\bf x})
\frac{d N_{>R}} {d \delta_c}.
\end{equation}
The large scale 
cluster-cluster correlation function is given by
$\xi_{CC}(x)= \left< \delta_C({\bf x}) \delta_C(0) \right>$ where
$x=|{\bf x}|$ and $\delta_C({\bf x}) = (N_{>R} ({\bf x}) - N_{>R})/
{N_{>R}}$ is the fractional overdensity in the cluster distribution. 
Defining bias $b$ via $\xi_{CC}(x)=b^2 \xi(x)$ 
where $\xi(x)=\left< \delta_b({\bf x}) \delta_b(0)\right>$  is the
correlation function of the dark matter on large scales, 
we find	
$
 b =
 1- (d
N_{>R} / {d \delta_c})/{ N_{>R}}$.

Substituting using equation~\ref{eqn-n} yields
the bias:
\begin{equation}
b=1+\frac{P[y_c(R)] / (R^{3} \sigma_R)
-3 \int_R^\infty dR'\; 
P[y_c(R')] / (R'^4 \sigma_{R'})}
{\int_R^\infty
dR'P[y_c(R')]y_c(R')\left(-d\sigma_{R'}/{dR'}\right)/
 (R'^{3} {\sigma_{R'}})}.
\end{equation}

For each of our models we
specify the background cosmology in terms of $\Omega_m$ and
$\Omega_\Lambda$. We then parameterize the 
power spectrum $P(k)$ of matter fluctuations as a CDM spectrum 
(Bardeen et al.\markcite{BBKS}~1986, eq.~G3) with shape
parameter $\Gamma$ (where $\Gamma \simeq \Omega h$) and
normalization specified by $\sigma_8$. Over the range of
scales relevant to cluster formation 
this form gives a useful fit to power spectra in
a variety of structure formation scenarios. We
choose the PDF $P(y)$ to be either a $\chi^2$ or a log-normal
distribution,
translated and renormalized to have mean zero and
standard deviation 1.
In each case there is one free parameter which
allows us to dial the amount of deviation from gaussianity, which 
we quantify in terms of a new parameter $T= 
\sqrt{2\pi}{\int_3^\infty P(y) dy}/{\int_3^\infty e^{-y^2/2} dy}
$, where
$T$ is the probability of obtaining a three or higher standard
deviation peak for the PDF in question relative to that for a 
gaussian PDF. To obtain log-normal and $\chi^2$
distributions with $T$ less than 1,
we reflect the PDF by making the transformation $P(y)\rightarrow P(-y)$.
Some sample PDFs, along with the
value of $T$, are illustrated in Fig.~1. 
Our models are thus
specified by five parameters: $\Omega_m$, $\Omega_\Lambda$, $\Gamma$,
$\sigma_8$ and $T$.

The PS formula is based on analytic results
for the collapse of spherically symmetric perturbations. In the case
of gaussian fluctuations, it provides remarkably good fits to the
multiplicity function of cluster scale objects (Gross et
al.\markcite{G98}~1997).
For  non-gaussian
fluctuations, one may worry that 
the spherical collapse picture is less applicable,
particularly in the case $T\ll 1$, where the
highest density objects will tend to form from matter swept 
out of 
low density regions. As for cluster correlations, 
our result
for the cluster bias in the gaussian case is identical to that of 
Mo \& White\markcite{MW}
(1996), which has been shown by Mo, Jing \& White\markcite{MJW} (1996)
to give a good fit to the amplitude of cluster correlations in N-body
simulations. For highly non-gaussian fluctuations, problems could arise in
the separation of the long and short
wavelength fluctuations, which are not necessarily
uncorrelated. However, if most clusters form from rare peaks in the
initial density field, and if the long and short wavelength modes are
nearly uncorrelated, then there is no reason to expect the modified PS
or peak-background split formalisms to lead to errors.
Our predictions seem to be born out qualitatively by the results
of published simulations of  non-linear  gravitational clustering 
from non-gaussian initial conditions (Park, Spergel \&
Turok\markcite{PST} 1991;
Weinberg \& Cole\markcite{WC} 1992; Borgani et al.\markcite{BCMP94}~1994).
However, a detailed quantitative
comparison with numerical simulations remains to be carried out. 

\section{Results}
\label{observations}
We compare our predictions with two sets of
observational data. Firstly, we compare predictions for the cluster temperature
function at $z=0.05$ with observations computed from the catalogue of
nearby clusters 
compiled by Henry \& Arnaud\markcite{HA} (1991).
We compute the cumulative number counts at two values of the
temperature, and estimate the covariance matrix by bootstrap 
resampling. For
each resampling, we add gaussian random errors to the temperature
consistent with the quoted observational uncertainty, and an
additional 10\% systematic error to allow for uncertainties in the $M$
vs. $T$ normalization (Eke, Navarro \& Frenk\markcite{ENF} 1998).
The observational data, along with $2\sigma$--errorbars for our two chosen
temperatures, are shown in the top panel of Fig.~2.

Secondly, we compare with the amplitude of the cluster-cluster correlation
function, quantified by $r_0$, the value of $r$ for which
$\xi_{CC}(r)=1$. Croft et al.\markcite{C97}~(1997) 
(hereafter C97) have computed $r_0$ as a function
of mean cluster separation $d$
for subsamples of the APM cluster survey with different lower bounds
on the cluster richness.  Their results are shown in
the bottom panel of Fig.~2. We compare model
predictions with just the $d=30h^{-1}$Mpc data point, since this point
is derived from the largest number of clusters, and the points are
highly correlated. We use $r_0=13.2 \pm 0.91 \pm 1.0 h^{-1}$Mpc,
where we have corrected the correlation amplitude for redshift space
distortions by subtracting $1h^{-1}$Mpc, 
as suggested by the simulations in \markcite{C97}C97.
The first errors are their $2\sigma$ statistical errors
and the second are $2\sigma$ errors we have
added to account for systematic uncertainties in 
the redshift space
correction.  
For each model we use the
techniques described above to compute the cluster correlation length
$r_0$ at $z=0$ for clusters with pre-collapse radius determined by
their mean separation 
$d=N_{>R}^{-1/3}=30h^{-1}$Mpc.   

In Fig.~2
we compare observations and predictions
for a set of four models, each with $\Omega_m=1$,
$\Omega_\Lambda=0$ and $\Gamma=0.2$.
Models 1, 2 and 3 have a normalization given by
$\sigma_8=0.5$. Model 
1 has gaussian fluctuations, model 2 has $T=0.0348$
(a reflected $\chi^2$ distribution with 50 degrees of
freedom), and model 3 has $T=16.3$
(a $\chi^2$ with 1
degree of freedom). 
Model 4 has the same PDF as model 3, but a
lower normalization, $\sigma_8=0.2$.

From the top panel of
Fig.~2
we see that
increasing $T$ increases the number
density of clusters of a given temperature. The reason for this is
clear from Fig.~1:
for the value of $\sigma_8$ considered here, 
only peaks of height roughly $3\sigma$ and higher (shown by the
heavily shaded region under each curve) will collapse, since 
the ratio of the critical overdensity to the standard deviation $\sigma_8$
in a sphere which will collapse to form a typical mass cluster is 
$y_c = \delta_c/\sigma_8=1.69/0.5 \simeq 3$.  So in this ``standard''
case, the parameter $T$
directly counts the expected number of clusters.

Now we consider the effect of changes in the PDF on the amplitude of
the cluster correlation function.
From the bottom panel of Fig.~2 we see that increasing
$T$ decreases the correlation amplitude, for 
reasons which are also illustrated in Fig.~1.
Adding a small background fluctuation 
modifies the
local value of the critical overdensity for collapse, leading to a
spatial modulation in the number density of collapsed objects.
The lightly shaded regions under the curves
in Fig.~1
show the effect for each PDF
of modifying $y_c$ from 3.0 to 2.5.
For
the $T>1$ PDF which has a long flat tail, this modification
produces only a small
enhancement in the local number of objects capable of collapse, and hence
there is only a small enhancement in the clustering amplitude of those
collapsed objects. For the $T<1$ PDF where the positive tail falls off very
sharply, this modification greatly
enhances the local number of collapsed objects, and consequently there
is
large enhancement in the clustering amplitude of those
objects.

Finally, we consider the effect  on each of the
above results of changing $\sigma_8$ for a fixed PDF.
Model 4 illustrates the effect of decreasing $\sigma_8$ for the case $T>1$.
As $\sigma_8$ is decreased, the number density of clusters decreases,
the reason being that 
the number of peaks with high enough over-density to collapse is
decreased. The top panel of Fig.~2 shows that for
$T=16.3$, $\sigma_8=0.2$ provides a 
good fit to the temperature function data. 
In the bottom panel of Fig.~2,
we see that as $\sigma_8$ is decreased, the 
amplitude of the cluster-cluster correlation function also
decreases. This is due to the fact that 
decreasing $\sigma_8$ decreases the amplitude of the underlying dark
matter correlation function, but leaves the bias roughly unchanged.
Model 4 shows that
for $T=16.3$, $\sigma_8=0.2$ provides a worse fit to the correlation
function data than $\sigma_8=0.5$.   
 
Thus we see that the two datasets we have chosen are complementary:
variations in the parameter $T$ introduce a degeneracy in the preferred
value of $\sigma_8$ for each dataset, but the degeneracies are
``orthogonal''. For the temperature function, increasing $T$ requires
lower $\sigma_8$, while for the correlation amplitude, increasing $T$
requires higher $\sigma_8$. So taken in combination, these
data
 allow a precise
determination of both $\sigma_8$ and $T$.
We quantify this by computing values of the $\chi^2$ statistic for each model
relative to the two datapoints in the top panel of Fig.~2 and the
$d=30h^{-1}$Mpc data 
point in the bottom panel.
In Fig.~3 we show goodness of fit contours in the
$\sigma_8$ vs.~$T$ plane for log-normal PDF models with a range of background
cosmologies and values of the parameter $\Gamma$. (We use three values of
$\Gamma$ which span the range allowed by measurements of the APM cluster
power spectrum by Tadros, Efstathiou \& Dalton\markcite{TED} 1997.) 

The dotted blue line ($T=1$) shows the location in the $T$
vs.~$\sigma_8$ plane 
of a gaussian PDF. 
The constraints on $\sigma_8$ from present day cluster number
abundances have 
been extensively investigated in this restricted case 
(White, Efstathiou \& Frenk 1993\markcite{WEF}; Eke et
al.~1996\markcite{ECF}; 
Viana \& Liddle\markcite{VL} 1996) and the results obtained
here are in close agreement.
In addition, Mo et al.\markcite{MJW} (1996) have
investigated the combined constraints of 
cluster correlation amplitude and number abundance measurements for
the case of gaussian fluctuations. Our results along this ``line of
gaussianity'' can be considered an update of their analysis, as
we use more recent cluster correlation data (C97)
with smaller observational errors. 

The results depend both on the
background
cosmology and the power spectrum shape parameter $\Gamma$.
For low values of
$\Omega_m$, the value of $\sigma_8$ preferred by the temperature
function data increases. This is just as
observed in the gaussian case,
and is due to the fact that for low values of
$\Omega_m$, clusters of typical mass form from 
spheres with  a
larger pre-collapse radius $R$, for which $\sigma_R$ is less than
$\sigma_8$.
The value of $\sigma_8$ preferred by the correlation
amplitude data, however, is almost independent of $\Omega_m$ and
$\Omega_\Lambda$. This is because the correlation amplitude is
measured for clusters with a fixed value of $d$, and consequently for
clusters with the same pre-collapse radius (not 
the same mass). Finally, as $\Gamma$ increases, 
the value of $\sigma_8$
preferred by the correlation amplitude also increases. This is because
the corresponding change in the shape of the power spectrum decreases 
the amount of power on scales
of order the cluster correlation length for a given value of
$\sigma_8$.

Results for the best fit values of $\sigma_8$ and $T$ from the
combined datasets for the family of log-normal PDFs in the various
background cosmologies are summarized in table~1. We have performed
the same calculation for $\chi^2$-parameterized 
PDFs and find very similar
results. Thus we expect that we can infer results for other models
provided we work out the 
appropriate value of the parameter $T$  
(for instance $T\simeq 14$ for textures -- Park et al. 1991). 
In future work we will expand
on the analysis presented here by including observations of the evolution of
cluster abundance and making use of updated datasets at low redshift.

\section{Conclusions}
\label{conclusions}

We have modified the Press-Schechter
and the peak-background split formalisms to compute
the cluster number density and cluster correlation
amplitude in non-gaussian models. The best-fit value of $\sigma_8$ and
level of non-gaussianity depend on the choice of background
cosmology, but we have demonstrated how these two quantities can be
constrained simultaneously. 
For a flat, $\Omega_\Lambda=0$, $\Gamma=0.2$ universe we find that the
fluctuations are consistent with gaussian, with
$\sigma_8=0.49^{+0.08}_{-0.07}$ (95\% limits). For low-$\Omega_m$
models with $\Gamma=0.2$, 
gaussian fluctuations are
ruled out with a high degree of confidence, while non-gaussian fluctuations
($T>1$) are strongly preferred. 
Taken in conjunction with
accurate knowledge of the background parameters (which we are rapidly 
gaining from
supernovae and CMB observations),
this method can provide a powerful constraint on the nature of
density fluctuations in the universe. 

Some concerns remain regarding the validity of 
our modification of the PS and peak-background split formalisms to the
case of non-gaussian fluctuations, and some systematic errors may not
yet be accounted for. This issue
should be settled by an analysis of n-body simulations with non-gaussian
initial conditions. However, our formulae have been well tested for
the gaussian case, so the result that gaussianity is ruled out for
certain choices of background model is not subject to these
uncertainties. At the very
least, we have demonstrated the power of two simple and
well measured datasets, the cluster temperature function and the
amplitude of cluster correlations, to simultaneously
constrain the amplitude of mass fluctuations and the level of
non-gaussianity in the universe.

We would like to thank P.~Ferreira, M.~Davis and M.~Markevitch for
helpful discussions. This work has been supported in part by a grant
from the NSF, and E.G. acknowledges the support of an NSF Graduate
Fellowship.

\begin{figure*}[t]
\centerline{\psfig{file=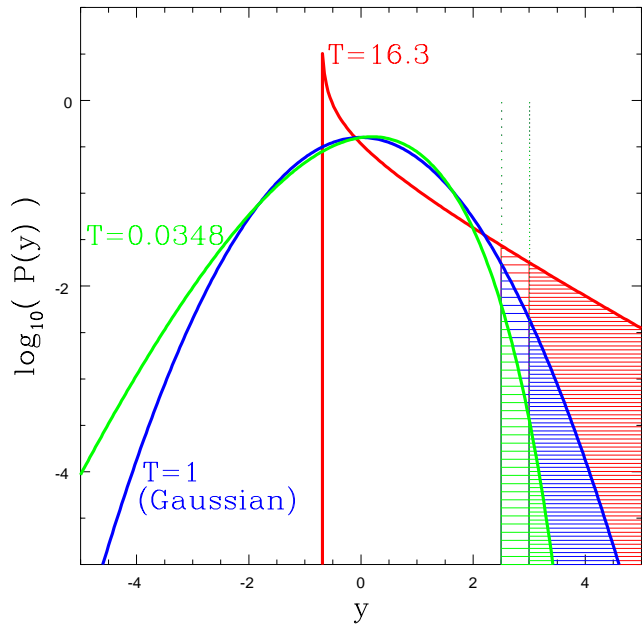,width=3.5in}}
\caption{
Three PDFs. The heavily shaded
region shows the area contributing to peaks of height $3\sigma$ and
above, while the lightly shaded region shows the additional
contribution to peaks of height $2.5\sigma$ and above.
}  
\label{fig-pdf}
\end{figure*}

\begin{figure*}
\centerline{\psfig{file=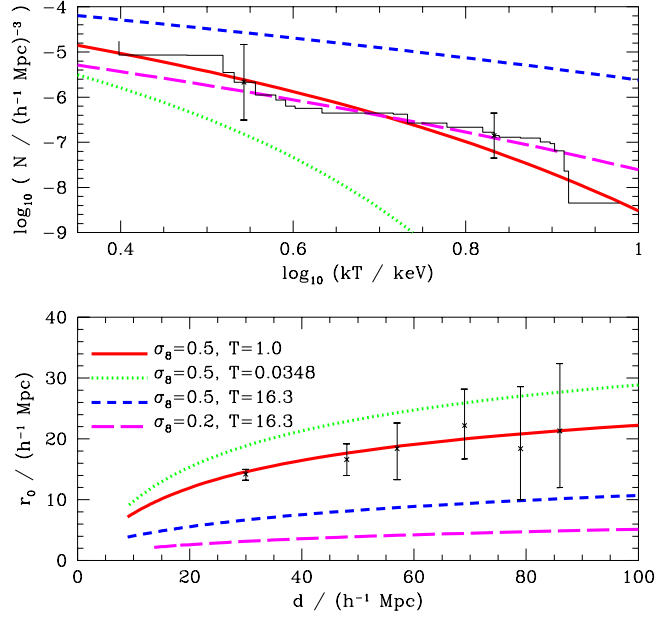,width=3.5in}}
\caption{
The cluster temperature function (black jagged
line, top panel), and the cluster correlation length $r_0$  
(datapoints, lower panel), with predictions for 
four
models ($\Omega_m=1.0$, $\Omega_\Lambda=0.0$, and $\Gamma=0.2$,
$\sigma_8$ and $T$ labeled in lower panel).
}
\label{fig-data}
\end{figure*}

\begin{figure*}
\centerline{\psfig{file=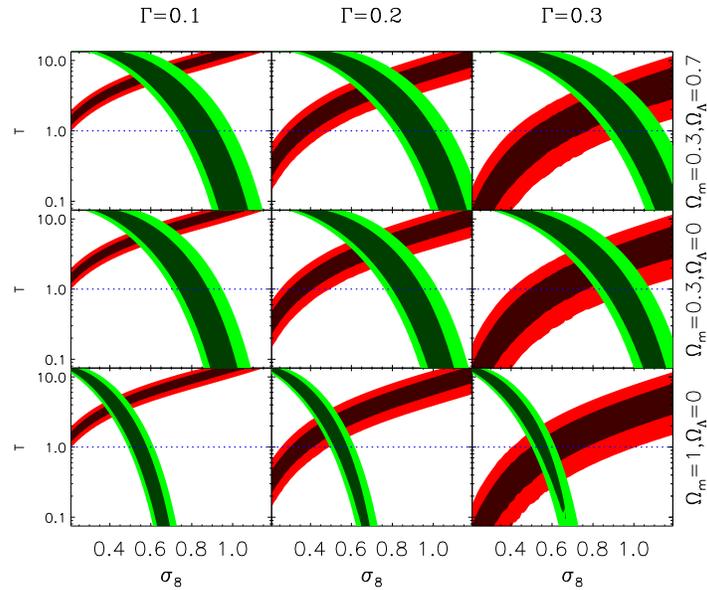,width=3.5in}}
\vskip 0.5in
\caption{
Confidence intervals (68\%---dark band, 95\%---light band) in the $T$
vs.~$\sigma_8$ plane 
for the cluster temperature function (green band, top left to bottom
right) and the cluster correlation length (red band, bottom left to
top right). The blue
dotted line corresponds to gaussianity ($T=1$), and the panels are
labeled with the values of
$\Omega_m$, $\Omega_\Lambda$, and $\Gamma$.
}  
\label{fig-pdf4}
\end{figure*}

\begin{table*}
\centering
\begin{tabular}{ccccc}
\tableline 
\tableline
$\Omega_m$& $\Omega_\Lambda$ & $\Gamma$ & $\sigma_8$  &      $T$  \\
\tableline
1.0      & 0.0            & 0.1    & $0.42^{+0.08}_{-0.05}$ & $3.8^{+1.7}_{-1.2}$ \\
1.0      & 0.0            & 0.2    & $0.49^{+0.08}_{-0.07}$ & $2.0^{+1.7}_{-1.1}$ \\
1.0      & 0.0            & 0.3    & $0.56^{+0.10}_{-0.08}$ & $0.9^{+1.3}_{-0.7}$ \\
0.3      & 0.0            & 0.1    & $0.59^{+0.15}_{-0.09}$ & $5.8^{+2.8}_{-1.8}$ \\
0.3      & 0.0            & 0.2    & $0.71^{+0.17}_{-0.13}$ & $3.8^{+3.2}_{-2.1}$ \\
0.3      & 0.0            & 0.3    & $0.83^{+0.21}_{-0.16}$ & $2.1^{+3.2}_{-1.6}$ \\
0.3      & 0.7            & 0.1    & $0.61^{+0.15}_{-0.10}$ & $6.3^{+3.1}_{-2.1}$ \\
0.3      & 0.7            & 0.2    & $0.73^{+0.19}_{-0.12}$ & $4.0^{+3.6}_{-2.0}$ \\
0.3      & 0.7            & 0.3    & $0.87^{+0.21}_{-0.17}$ & $2.3^{+3.5}_{-1.7}$ \\
\tableline
\end{tabular}
\vskip 10pt
\caption{
Best fit values of $\sigma_8$ and $T$, with 95\% confidence limits.
}
\label{tab}
\end{table*}

\end{document}